# Tuning magnetic anisotropy in (001) oriented $L1_0$ (Fe$_{1-x}$Cu$_x$)$_{55}$Pt$_{45}$ films

Dustin A. Gilbert[1], Liang-Wei Wang[2], Timothy J. Klemmer[3], Jan-Ulrich Thiele[3],

Chih-Huang Lai[2], and Kai Liu[1,*]

[1]*Physics Department, University of California, Davis, CA 95616 USA*

[2]*Department of Materials Science and Engineering, National Tsing Hua University, Hsinchu, Taiwan*

[3]*Seagate Technology, Fremont, CA 94538 USA*

**Abstract**

We have achieved (001) oriented $L1_0$ (Fe$_{1-x}$Cu$_x$)$_{55}$Pt$_{45}$ thin films, with magnetic anisotropy up to $3.6 \times 10^7$ erg/cm$^3$, using atomic-scale multilayer sputtering and post annealing at 400 °C for 10 seconds. By fixing the Pt concentration, structure and magnetic properties are systematically tuned by the Cu addition. Increasing Cu content results in an increase in the tetragonal distortion of the $L1_0$ phase, significant changes to the film microstructure, and lowering of the saturation magnetization and anisotropy. The relatively convenient synthesis conditions, along with the tunable magnetic properties, make such materials highly desirable for future magnetic recording technologies.

The development of high anisotropy magnetic materials that are compatible with industrial processing is critical in advancing magnetic recording,[1-5] permanent magnet,[6, 7] and spintronic technologies.[8, 9] Specifically, high anisotropy materials are necessary to ensure long-term thermal stability in magnetic nanoelements, such as ultra-high density recording media and magnetic memory.[10] A material of particular interest is $L1_0$ ordered FePt because of its large magneto-crystalline anisotropy ($K_U$), saturation magnetization ($M_S$), and chemical stability (compared to alternatives e.g. SmCo).[11] A key limiting factor has been the high annealing temperature necessary to transform the as-deposited disordered face centered cubic (fcc) $A1$ phase into the ordered tetragonal $L1_0$ phase. Prospects of creating FePt-based ternary alloys have been studied to lower the annealing temperature and to improve the orientation and granular structure.[12-16] In particular, it has been suggested that the inclusion of Cu lowers the kinetic ordering temperature, which is highly desirable.[2-5] The consequences of adding Cu to FePt has often been explored by forming $(FePt)_{1-x}Cu_x$ [along ray (*i*) of the phase diagram[17] shown in Fig. 1(a)]. It is generally accepted that Cu replaces Fe lattice sites in FePt,[18] thus this approach quickly deviates from an equiatomic ratio of the A(Fe or Cu):B(Pt) site atoms ideal for the $L1_0$ lattice. Additionally, it has been challenging to realize FeCuPt films with strong (001) orientation, and to quantitatively evaluate the uniaxial magneto-crystalline anisotropy.

In this work, we have achieved (001) orientated $L1_0$ $(Fe_{1-x}Cu_x)_{55}Pt_{45}$ thin films using atomic-scale multilayer sputtering (AMS) and rapid thermal annealing (RTA) at 400°C for only 10 s, which is quite benign compared to earlier studies.[2, 18, 19] The composition of Pt was kept constant, so that the magnetic properties of the films, particularly the anisotropy, were systemically tuned along ray (*ii*) of the phase diagram in Fig. 1(a). Quantitative evaluations showed that the magnetic properties of these films were desirable and tunable for future magnetic recording applications, particularly as energy-assisted recording media.

Atomic scale multilayer films of (Fe/Cu/Pt)$_{16}$ were grown by DC magnetron sputtering from elemental targets on amorphous SiO$_2$(200 nm)/Si substrates in a vacuum chamber with base pressure of $7\times10^{-7}$ torr.[20] The multilayer structure with extremely thin layers establishes an artificial ordering which is conducive to the formation of (001) oriented tetragonal $L1_0$ phase upon annealing.[21, 22] By varying layer thicknesses the stoichiometric ratio was systematically tuned to achieve (Fe$_{1-x}$Cu$_x$)$_{55}$Pt$_{45}$, where the Cu content is adjusted to 0, 2, 5, 8, 11, 16, 22, and 27 at.%. The platinum thickness was fixed at 1.4 Å (1 ML), while Fe and Cu thicknesses were changed to the desired Fe:Cu:Pt ratio. The total film thickness was ~ 45 Å. The films were then annealed by RTA at 400°C (10s rise and 10s dwell time) in $7\times10^{-7}$ torr vacuum.

X-ray diffraction (XRD) with Cu K$_\alpha$ radiation was used for phase identification and lattice parameter determination. Atomic/magnetic force microscopy (AFM/MFM) was performed using a Si cantilever with a 15nm (magnetic) CoCr coating. Magnetic measurements were performed using vibrating sample magnetometry (VSM) and superconducting quantum interference device (SQUID) magnetometry in the out-of-plane configuration at room temperature, unless noted otherwise.

The as-sputtered films are polycrystalline, in the disordered cubic $A1$ phase (not shown). After annealing, the films convert into the tetragonal $L1_0$ phase and orient along the originally forbidden (001), as shown in Fig. 1(b). Following procedures outlined previously,[23-25] the order parameter, $S = (I_{(001)} + I_{(003)})/2I_{(002)}$, is calculated to gauge the degree of $L1_0$ ordering. Here $I_{(001)}$, $I_{(002)}$, and $I_{(003)}$ are the integrated intensity of the respective peaks, corrected for the Lorentz, Debye-Waller, and angular-dependent atomic scattering factors.[23, 26] The order parameter is normalized to the calculated maximum value, $S_{Max}=1-2\Delta$, where $\Delta=0.05$ is the variation in the elemental ratio from 50:50. Large values of $S/S_{Max}$ close to 1, shown in Fig. 1(c), are found for all the post-annealed films, confirming the presence of a predominantly $L1_0$ phase. The realization of (001) oriented $L1_0$ (FeCu)Pt under a significantly reduced thermal cycle highlights the advantage of the AMS growth.

Furthermore, the shifting peak positions with Cu content in Fig. 1(b) indicate a continuous change of the lattice parameters in the $L1_0$ films. Additional θ-2θ scans were performed at Ψ=54.7° to probe the (111) peak. The extracted lattice parameters are shown in Fig. 1(d). The unannealed cubic FePt film (0 at.% Cu) displays a lattice parameter of 3.80 Å. With increased Cu, the cubic lattice parameter remains largely unchanged. Upon annealing, the cubic lattice of the $A1$ phase shrinks in the $c$-axis and expands slightly in the $a$-$b$ plane, forming the tetragonal $L1_0$ phase.[16] In FePt, the lattice parameters are $c$=3.72 Å and $a$=3.83 Å. The addition of Cu increases $a$ and decreases $c$, hence increasing the tetragonal distortion of the cubic lattice as indicated by the $c/a$ ratio, consistent with the gradual replacement of Fe by the smaller Cu atoms. The values are in agreement with previously reported results.[4]

Topographical information of the post-annealed films is shown in Figs. 2(a, c, e, g), obtained by AFM. The films have dispersed voids, increasing in size and density with increased Cu content. In fact, films with 27 at.% Cu form isolated islands with an average diameter of 36 nm [Fig. 2(g)]. Interestingly, the islands retain their (001) orientation – as shown by XRD in Fig. 1(b). Island formation in similar FePt ultrathin films has been previously observed and can be attributed to surface energy and lattice strain.[27, 28] Corresponding magnetic domain structures imaged by MFM over the same areas are shown in Figs. 2(b, d, f, h), after ac demagnetization in the out-of-plane direction. Large domains (~ 500nm) exist in the continuous FePt film, Fig. 2(b).

As the Cu content is increased the emerged voids act as pinning sites, deforming the previously smooth domain structure, Figs. 2(d, f). For the sample with 27 at.% Cu, Fig. 2(h) reveals that the isolated islands form single domains, with up and down moments (black and white contrasts), indicating that the exchange interaction between islands is greatly suppressed.

Representative magnetic hysteresis loops for the $L1_0$ films are shown in Figs. 3(a-d) with out-of-plane (solid symbols) and in-plane (open symbols) magnetic field directions. Coercivity (out-of-plane) and normalized remanence values (both in-plane and out-of-plane) for all the films are shown in Fig. 3(e). The loops all show a high out-of-plane, and a low in-plane, remanence confirming a strong out-of-plane (perpendicular) anisotropy. With increased Cu content the coercivity ($H_C$) initially decreases, which can be attributed to the decrease in the anisotropy. However, at high Cu ratios $H_C$ is shown to then increase, corresponding to the increased pinning sites (and eventual isolation) resulting from the microstructure.

Anisotropy values were found by evaluating the hard axis saturation field, or anisotropy field $H_K$, according to $K_U = M_S(H_K + 4\pi M_S)/2$. As shown in Fig. 4(a), $M_S$ decreases rapidly with increased Cu, from 1095 emu/cm$^3$ for 0 at.% Cu, consistent with previously published values of 1125-1280 emu/cm$^3$,[11, 12, 29] to 491 emu/cm$^3$ for 27 at.% Cu. These values are much larger than those of CoCrPt alloys (up to 410 emu/cm$^3$) currently used in perpendicular recording media.[1] Both $H_K$ and $K_U$ also decrease with increased at.% Cu, with $H_K$ ranging from 54 kOe to 26 kOe and $K_U$ from 3.6×10$^7$ erg/cm$^3$ to 8×10$^6$ erg/cm$^3$ [Fig. 4(b)]. These values in general agree with the theoretical calculations by Suzuki et al.[30] They are significantly larger than those of the CoCrPt alloy ($K_U$ between 2×10$^6$ and 2×10$^7$ erg/cm$^3$).[1, 11, 19]

For comparison, 50nm thick films of $(Fe_{1-x}Cu_x)_{55}Pt_{45}$ were grown by co-sputtering from composite targets of $Fe_{55}Pt_{45}$ and $Cu_{55}Pt_{45}$ onto glass substrates, with a 20nm MgO capping layer.

The composition of the films was tuned also along ray (ii) of Fig. 1(a). The resultant films were polycrystalline, both before and after annealing in vacuum ($3\times10^{-8}$ torr) at 700 °C for 1hr. Structurally, these films have similar lattice parameters as the AMS grown ones (within 0.03Å) and exhibit the same trend of increasing $a$ and decreasing $c$ with increased Cu content. The voids observed in the thinner AMS grown thin films are absent in these thicker films. The $K_U$ values could not be measured due to the polycrystalline nature of the films. Curie temperature ($T_C$) shows a continuous drop from 695 K (0 at.% Cu), similar to previously reported values of 730-770 K,[4, 12, 31] to 327 K (28 at.% Cu) in an almost linear fashion as shown in Fig. 5, similar to that found in $(Fe_{1-x}Ni_x)_{55}Pt_{45}$.[12] Interestingly, comparing disordered (unannealed) films to ordered (annealed) films of equal composition, $T_C$ increases with ordering for low Cu content while it decreases for high Cu concentrations. These values of $T_C$ are promising for energy assisted recording, where a moderately lower $T_C$ allows faster, lower energy writing.[32]

Previously Mryasov showed that magnetic properties of ordered $L1_0$ FePt alloys are determined by (1) "direct" Fe-Fe interactions between an Fe atom and its four nearest neighbors in the Fe layer and (2) polarization driven "indirect" Fe-Pt-Fe interactions.[33] By transitioning from the disordered to the ordered state, the formation of Fe planes guarantees that each Fe atom has four nearest magnetic neighbors, which improves the Fe-Fe interaction relative to the disordered state. However, as Cu addition replaces Fe, in the ordered state each Fe atom has fewer magnetic nearest neighbors (as low as two for the sample with the highest Cu content), reducing the "direct" interaction significantly. A further decrease in the "indirect" interactions is also experienced as the Fe is coupled to nonmagnetic Cu through the polarization of the Pt in the adjacent layer. The suppression of these interactions leads to the significant decrease in $T_C$ with Cu content.

The effects of Cu addition in FePt are shown in Fig. 6, compared with Ni[4, 12] and Mn[29, 31] based ternary alloys, plotted as a function of the effective electron density ($n_{eff}$). The $n_{eff}$ is determined by the valence electrons on the Mn (7), Fe (8), Ni (10), and Cu (11) atoms, using a linear interpolation with the respective concentrations.[30] The comparison shows that FeCuPt behaves similarly as FeNiPt, and closely reflects the expected trend.[30] Copper inclusion is thus shown as an effective tool to gradually tune (compared to e.g. Mn) the magnetic properties over a large parameter space. This, coupled with the reduced kinetic ordering temperature suggested previously,[2-5] makes FeCuPt an attractive candidate for applications in magnetic recording and spintronics. Also, the superior thermal conductivity of Cu suggests that heat transport, particularly across the (001) planes through Cu-Pt pathways, should be improved compared to other ternary alloys. This allows faster heat removal from the bit post writing, thus increasing writing speeds and reducing lateral thermal spread.[19]

In summary, we have achieved (001) oriented $L1_0$ $(Fe_{1-x}Cu_x)_{55}Pt_{45}$ thin films using AMS and RTA at a much lower annealing temperature and shorter annealing time. By keeping the Pt concentration constant, the effect of Cu-alloying on the magnetic properties of the films is systematically investigated. In the annealed FeCuPt thin films, the increased Cu content leads to a progressively larger tetragonal lattice distortion, consistent with the replacement of the Fe atoms by the smaller Cu atoms. Also, in the thinner films grown by AMS, annealing caused the partial de-wetting of the film resulting in voids with striking differences in morphology dependent on the Cu-content of the films. For large Cu content the film became discontinuous, instead forming magnetically isolated islands, while still retaining a high magnetic anisotropy, crystallographic orientation, and chemical ordering. As expected, the saturation magnetization and magnetic anisotropy both decrease with increasing Cu content. However they are still substantially larger than the CoCrPt alloys currently in use. A second series of $(Fe_{1-x}Cu_x)_{55}Pt_{45}$ films made by co-sputtering further shows that the $T_C$ can be continuously tuned with Cu content. These results demonstrate a relatively simple approach to achieve high quality $L1_0$ FeCuPt films that are desirable for future magnetic recording technologies, and the magnetic properties can be smoothly tuned by Cu-substitution into select sites of the ordered alloy.

This work has been supported by the NSF (DMR-1008791). Work at NTHU has been supported in part by Hsinchu Science Park of Republic of China under Grant No. 101A16.

**Figure Captions**

**Fig. 1:** (Color online) (a) Phase diagram for ternary alloy FePtCu determined at 600° C (Values shown are at.%, adapted from Ref. 17). Sample compositions in this study are indicated by crosses. (b) XRD patterns and (c) normalized order parameter $S/S_{Max}$ for post-annealed films grown by AMS. The at.% Cu in (b) is 0, 5, 11, and 22 from top to bottom, respectively. Error bars in (c) are determined by the standard deviation of $I_{(001)}/I_{(002)}$ and $I_{(003)}/I_{(002)}$. (d) Lattice parameter $a$ for the $A1$ phase (squares), $a$ (solid circles) and $c$ (open circles) for the $L1_0$ phase, as well as the $c/a$ ratio for the latter (triangles).

**Fig. 2**: AFM (top row) and corresponding MFM (bottom row) images for $L1_0$ ordered films with (a, b) 0, (c, d) 8, (e, f) 16, and (g, h) 27 at.% Cu. The scale bar indicates 500nm.

**Fig. 3:** (Color online) Hysteresis loops of AMS grown $L1_0$ films with (a) 0, (b) 8, (c) 16, and (d) 27 at.% Cu (same films as in Fig. 2). The dependence of remanence and coercivity (triangles) on the Cu content is shown in (e). Solid squares and open circles represent the out-of-plane and in-plane field direction, respectively.

**Fig. 4**: (Color online) Dependence of (a) saturation magnetization $M_S$, (b) anisotropy constant $K_U$ and anisotropy field $H_K$ for AMS grown $L1_0$ $(Fe_{1-x}Cu_x)_{55}Pt_{45}$ films as a function of Cu content. The results in general agree with theoretical values from Ref. 30.

**Fig. 5**: (Color online) Curie temperature variation with Cu content in co-sputtered $(Fe_{1-x}Cu_x)_{55}Pt_{45}$ films, before and after annealing at 700°C for 1 hr.

**Fig. 6**: (Color online) (a) Saturation magnetization, (b) uniaxial anisotropy, and (c) Curie temperature variation with ternary element content for alloys of FePtCu [AMS samples in (a) and (b), co-sputtered samples in (c), Ref. 4 is also included in (c)], FePtNi (Ref. 12, 31) and FePtMn (Ref. 29, 31). Dotted lines in (a, b) indicate calculated values (Ref. 30).

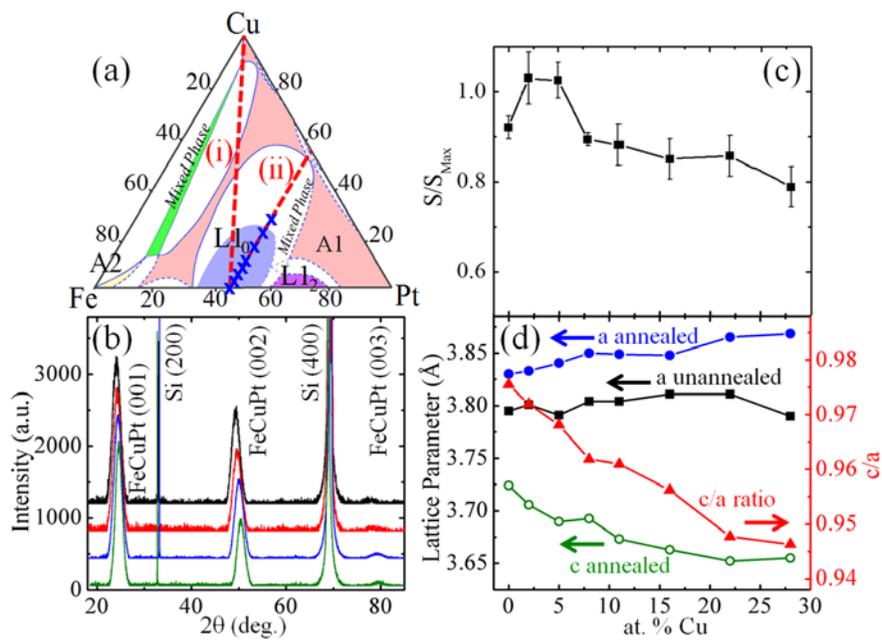

Figure 1

Gilbert et al.

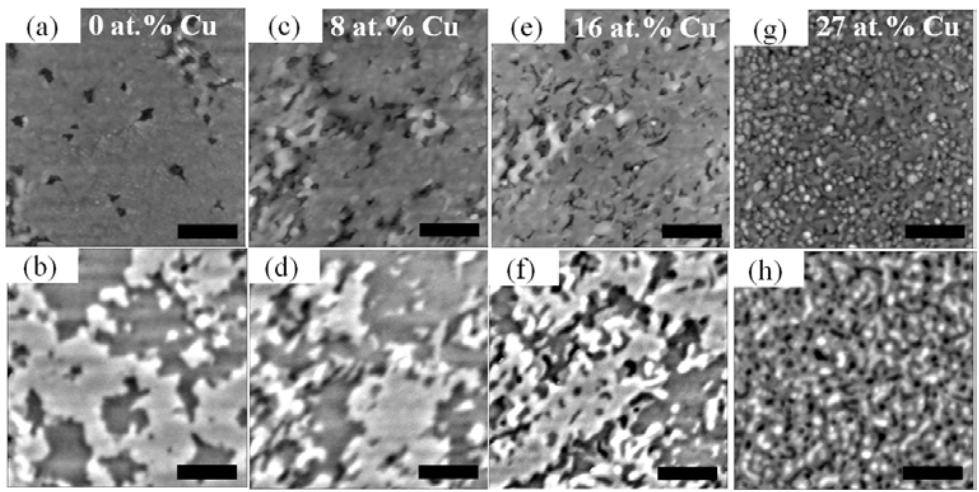

Figure 2

Gilbert et al.

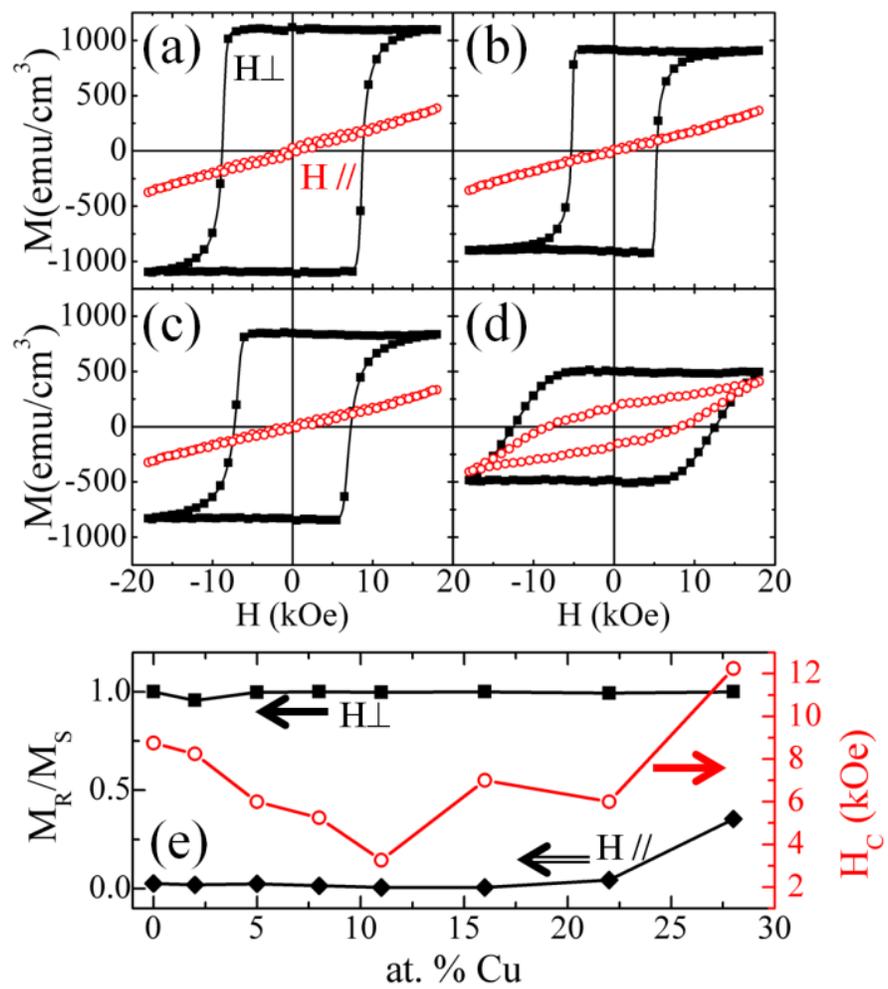

Figure 3

Gilbert et. al

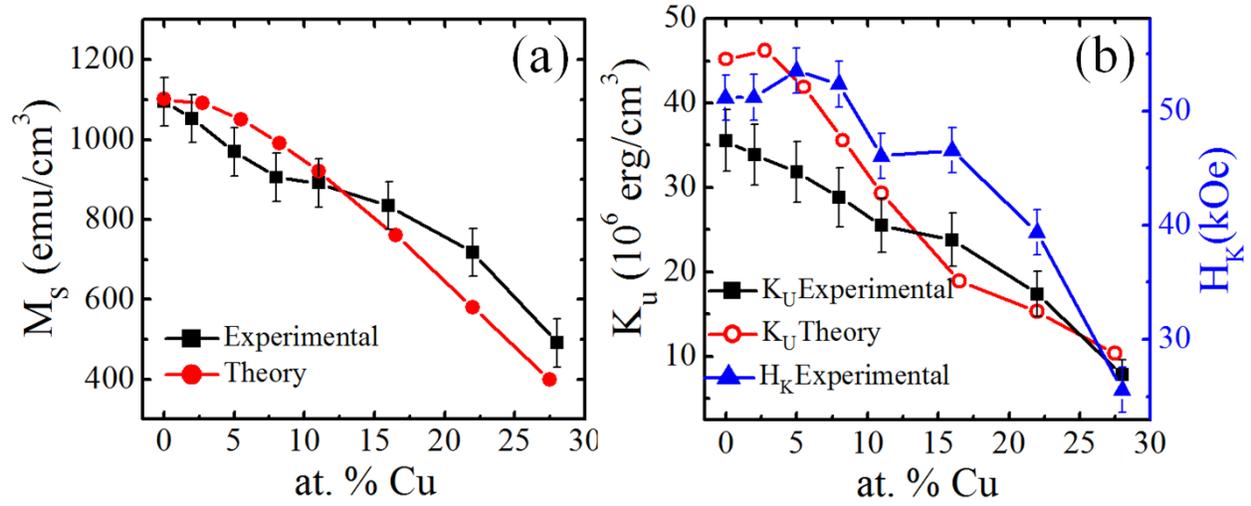

Figure 4

Gilbert et al.

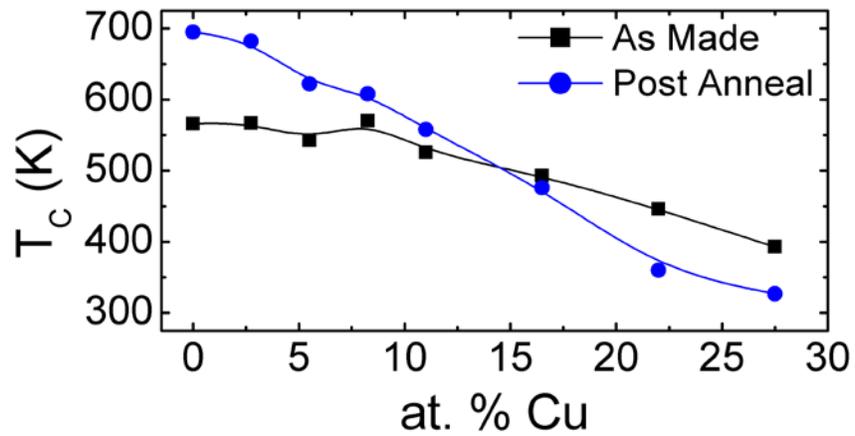

Figure 5

Gilbert et al.

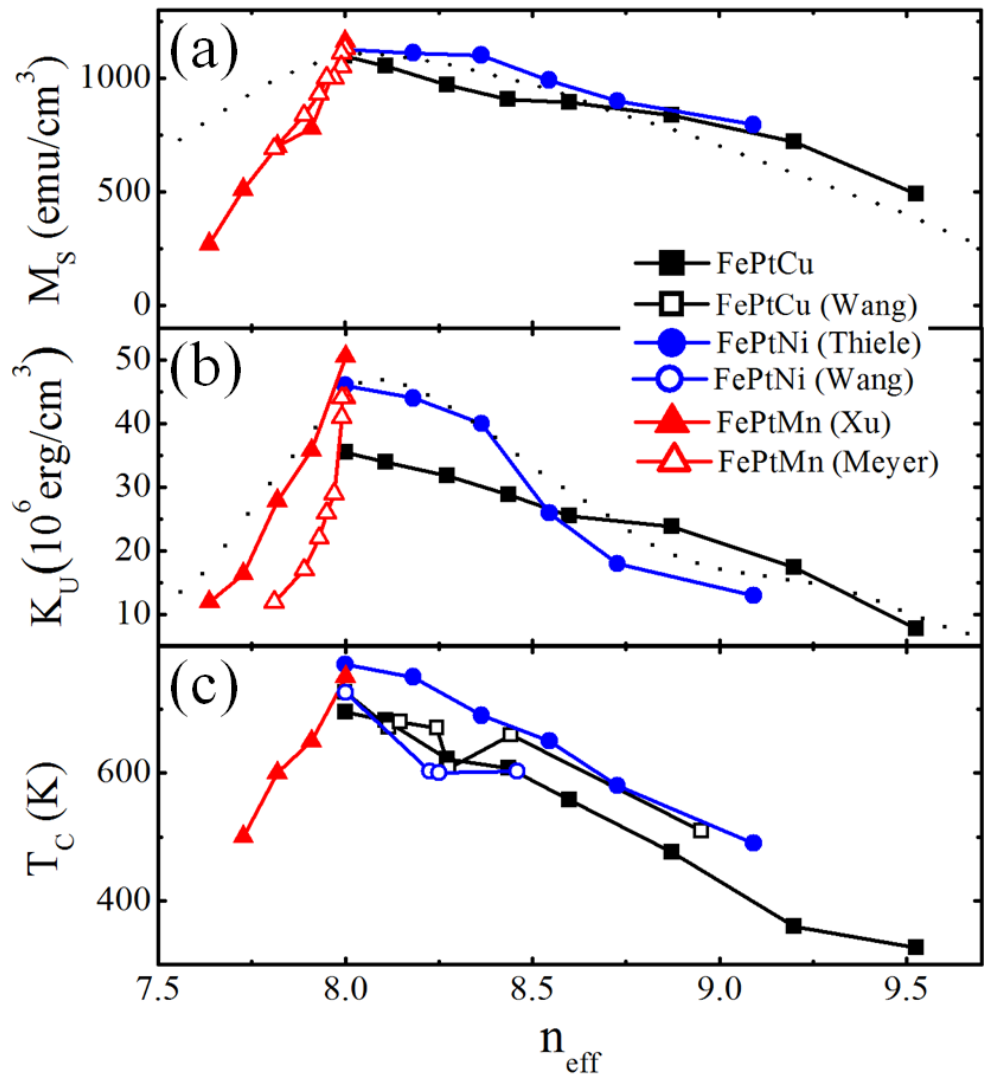

Figure 6

Gilbert et al.